\RequirePackage{fix-cm}
\documentclass[smallextended]{svjour3} % onecolumn (second format)
\smartqed % flush right qed marks, e.g. at end of proof
\usepackage{graphicx}

\usepackage{mathptmx} % use Times fonts if available on your TeX system

\usepackage{times}
\usepackage{bm} % for math
\usepackage{amssymb} % for math
\usepackage{amsmath} % for math
\usepackage{braket} %for brakets
\usepackage{mathrsfs} %for script letters 

\begin{document}
%%%FrontMatter%%%%%FrontMatter%%%%%%%%%FrontMatter%%%%FrontMatter%%%%%%%%%%FrontMatter%%%
\title{Characterizing the Nash equilibria of a three-player Bayesian quantum game}

\author{Neal Solmeyer \and Ricky Dixon \and Radhakrishnan Balu }
%\authorrunning{Short form of author list} % if too long for running head}
\institute{N. Solmeyer \at
Sensors and Electron Devices Directorate, Army Research Laboratory, Adelphi, MD, 21005-5069, USA. \email{neal.e.solmeyer.civ@mail.mil} 
\and 
R. Dixon \at
Mississippi Valley State University, 14000 Hihghway 82 West, Itta Bena, MS, 38941.
\and
R. Balu \at
Computer and Information Sciences Directorate, Army Research Laboratory, Adelphi, MD, 21005-5069, USA. 
\and 
}
\date{Received: date / Accepted: date}

\maketitle

\begin{abstract}
Quantum games with incomplete information can be studied within a Bayesian framework. We consider a version of prisoner's dilemma (PD) in this framework with three players and characterize the Nash equilibria. A variation of the standard PD game is set up with two types of the second prisoner and the first prisoner plays with them with probability $p$ and $1-p$ respectively. The Bayesian nature of the game manifests in the uncertainty that the first prisoner faces about his opponent's type which is encoded either in a classical probability or in the amplitudes of a wave function. Here, we consider scenarios with asymmetric payoffs between the first and second prisoner for different values of the probability, $p$, and the entanglement. Our results indicate a class of Nash equilibria (NE) with rich structures, characterized by a phase relationship on the strategies of the players. The rich structure that can be exploited by the referee to set up rules of the game to push the players towards a specific class of NE. These results provide a deeper insight into the quantum advantages of Bayesian games over their classical counterpart. 
\end{abstract}

%%%Body%%%%%Body%%%%%%%%%Body%%%%Body%%%%%%%%%%Body%%%
\section{Introduction}

Game theory as a sub field in mathematics \cite{Math} has enjoyed a tremendous growth and has been applied to wide range of fields such as economics \cite{Econ}, political science \cite{Pol}, biology \cite{Bio}, and computer science \cite{CS}. The successful application of the theory in the classical context inspired formulation of quantum games \cite{Meyer}. One key feature of a quantum game is that there are an infinite number of strategies possible, which can potentially lead to far more equilibria. In quantum game theory, concepts from quantum information theory are applied to game theory such that qubits represent the states of each player, quantum gates (unitaries) are used for implementing strategies, and entanglement is used to mediate communication between the players. Considerable interest was generated in this field when the quantization of a classical game using the Eisert-Wilkens-Lewenstein (EWL) formalism \cite{EWL} showed that the classical prisoner's dilemma (PD) could be resolved by including a new quantum strategy that is not available in the classical game. We employ the EWL formalism primarily because it is widely used in the literature. This formalism quantizes the strategy space of a classical game as opposed to other approaches where the payoff function is quantized \cite{Bleiler}, which can be advantageous for directly comparing a quantum game to mixed strategy classical games. Extension of a classical game into a quantum context gives rise to entirely new classes of games, depending on how that extinsion is made, and our focus here is again on games based on EWL formalism. There is also distinction between quantizing a classical game and gaming a quantum system \cite{Travis} as the former is concerned with applying quantum information to game theory, with the goal to learn something about the game that is produced, whereas the latter applies the rules of game theory to quantum physics, with the goal of learning something about the underlying physics. Our approach is the former. 

In this work we are focused on quantizing classical Bayesian games that have players with incomplete knowledge about the payoff functions of their opponents. The uncertainty in knowledge of the players is encoded in types and priors or beliefs as classical probability distributions on types. Formally, a classical two person Bayesian game is a tuple $(\Omega\times\Omega, \rho\times\rho, \mathbb{A}\times\mathbb{A},\mathbb{X},\mathbb{F})$ where, $\Omega, \mathbb{A}$ represent the state and action spaces of the player, $\mathbb{X}$ is space of types from which nature assigns one member for each player. Lack of knowledge on the types makes this space a random variable and each player has priors or beliefs about it in the form of a probability distribution on it which is encoded in the probability measure $\rho$. This results in players choosing a strategy from $\mathbb{A}$ conditioned on the types and the cost function $\mathbb{F}$ is a mapping on $\mathbb{A}\times\mathbb{X}\rightarrow\mathbb{R}$. Later, we describe the quantum version of this Bayesian set up, where the priors are still classical but  $\Omega$ and $\mathbb{A}$  are based on quantum information. Our approach is similar in spirit to the quantized Bayesian game of battle of sexes \cite{iqbal} where the probabilities are calculated in accordance with quantum mechanics and the priors are classical, however, their approach relies only on probability distributions, our approach retains the quantum mechanical formalisms of state vectors and operators. 

The primary advantage a quantum game has over its classical counterpart may be seen in the case of PD. When the initial state is maximally entangled, and the strategy space is restricted to those in the original (EWL) formalism \cite{EWL}, the payoff for the players at the Nash equilibrium (NE) in the quantum game exceeds the payoff of the players at the NE in the classical counterpart. The NE is the set of strategies where no player can benefit by unilaterally changing their strategy. It is possible in the classical game for the prisoners to choose strategies corresponding to a Pareto efficient (PE) solution which gives higher payoff than the NE, though this is only with cooperation or a contract, which require communication and can be broken. A PE solution is a set of strategies where no player can benefit without, unilaterally or not, hurting the other.   

A second key feature of quantum games is that the player's initial states can be entangled. Though we only look at non-cooperative games where the players act rationally and only in their self interest, the entanglement can ensure that the outcome of the player's strategy choices are correlated in a quantum mechanical way. Once established, these correlations persist even if the players cannot communicate, exhibiting the non-local characteristic of quantum mechanics. In some cases, the correlations produced from entanglement cannot be described classically. The role of entanglement in a quantum game has been interpreted as a form of advice, contract, or mediated communication between the players \cite{Benjamin2001b}, and is given by the referee. However, in contrast to classical game theory, the advice or contract is established before the players make their strategy choices by using an initial entangled state, after which, there is no communication between the players, and they are physically prohibited from breaking the contract. Entanglement can also be thought of as an environment that acts to correlate the player's choices, since the entanglement is imposed by the referee. 

The solutions to the game can vary greatly with the amount of entanglement such that new NE form with partial entanglement that are different from those at maximal entanglement or zero entanglement, or can also lead to the complete absence of a NE for an entangled game even when the classical game has a NE, such as in the maximally entangled PD game\cite{Landsburg}. Games with mixed strategies, that is, when the players choose the strategies with a probability, always have NE solutions, but may not have mixed strategy solutions for every possible probability distribution on the payoff functions. In other words there are distributions over the payoff functions that are not in the image of the mixed version of the original clasical game. One possible way to extend the games to realize other probability distributions over the payoff functions is to set up games where the strategies of the players are correlated by some form of advice or mediated communication. On the other hand, quantum games can realize every possible probability distribution on payoff functions through the use of entanglement which can facilitate correlated strategies. In fact, given one player's strategy the other player can choose strategies such that any possible distribution on the payoff function can be achieved\cite{Bleiler}, which can lead to the absense of NE in certain cases, such as in the maximally entangled PD game.

When there is incomplete information available to the players involved, the game can be treated using a Bayesian approach \cite{Noah}. This produces a game that is a classical mixture of two quantum games. Bayesian games have seen interest because they can be easily formulated to show a quantum advantage. This can be done by leveraging Bell's inequalities such that the payoff function of the game is cast in terms of the expectation values of observables employed in a form of a Bell's inequality. Thus, by using quantum correlations, a higher payoff at the NE can be achieved than is possible using only classical correlations. This shows the advantage for quantum games when the payoff function has the form of a Bell's inequality \cite{Situ2016,Iqbal2015}. We wish to study Bayesian games in a more general framework in order to shed light on how the non-local advice via the entanglement, functions in a game with incomplete information. 

The ability to include multiple agents, allowance for incomplete information, and the incorporation of game theory concepts such as fairness or equilibria make quantum Bayesian games a useful tool for quantum network, which features entangled qubits shared non-locally across multiple nodes. When interacting with multiple agents on a network, a game theory analysis is often justified because agents are typically free to make their choices in their best interests, perhaps mitigated by some referee. It is also often the case in interacting with multiple agents on a network, that the players will have incomplete information about the other agents, which rationalizes the approach to include a classical Bayesian framework for prior information on a quantum game. The utilization of quantum games on a quantum network could potentially be used in applications such as the analysis of the quantum security protocols\cite{Maitra2015}, the development of distributed quantum computing algorithms\cite{Li2009}, or using non-locally shared quantum information to improve the efficiency of classical network algorithms \cite{Zabaleta2017}. In addition there have been several experimental implementations of the two-player PD game within this framework using nuclear magnetic resonance \cite{Du2002}, quantum circuits in optical \cite{Zeilinger}, and ion-trap platforms \cite{Shuichi}.

The structure of the paper is as follows. In Section \ref{sec:classical} we introduce the classical game we wish to quantize and summarize its solutions. In Section \ref{sec:quantum} we give some of the theoretical background necessary to employ a Bayesian game within a quantum probability space. In Section \ref{sec:1} we give details of the method we employ to find the solutions to the quantized game, where we vary the degree of entanglement and amount of incomplete information. In Section \ref{sec:2} we present solutions to the two-player games where we find that the structure of each NE is comprised of a class of strategy choices related to one another by a phase relationship, giving rise to a class of NE between the two players. In Section \ref{sec:3} we analyze the quantum Bayesian game and find that the NE form a phase-diagram like structure in the amount of entanglement, and amount of incomplete information. The NE are found to have a complex and sometimes surprising structure within this phase diagram. Finally we offer a discussion of what our results tell us about the role of entanglement and partial information in a quantum Bayesian game in Section \ref{sec:4}, and end with some conclusions.

\section{Classical game background}
\label{sec:classical}

The Bayesian game we considered here is a variation of PD, the District Attorney’s (DA's) brother, which involves three players, or equivalently, represents a situation where player B does not know what type player A is. In the first case, player A and B play the standard PD game. In the second case, player B believes player A is the DA's brother which gives player A an advantage resulting in an asymmetric payoff between the players. 

First, we consider a version of the two-player PD game that is slightly modified from the canonical formulation such that the players have asymmetric payoffs. This has the payoff matrix given by:

\begin{center}
\begin{tabular}{ l || c | r }
$A| B_1$ & $ \Ket{0}(C)$ &$\Ket{1}(D)$\\
\hline 
\hline
$\Ket{0}(C)$& $ (11,9)$ & $(1,10)$ \\
$\Ket{1}(D)$& $ (10,1)$ & $(6,6)$ 
\end{tabular}
\end{center}

The payoff is computed for example, if $A$'s qubit is measured to be in $\Ket{0}$ and $B_1$'s is measured in $\Ket{1}$, then player A receives a payoff of 1, and player $B_1$ receives a payoff of 10. 

The interpretation of the classical game is that two prisoners, $A$ and $B_1$, are given the choice to implicate the other player in a crime, or to remain silent. Their classical strategy choices, are to defect or cooperate (implicate/remain silent), which we label D/C, with their payoff (i.e. jail sentence) determined by their joint choices. A higher payoff represents less time in jail. If they both play C, then neither admits to the crime, and their sentence is light, i.e. payoffs of 11 and 9 in our example. If they both implicate the other in the crime, that is, play strategy D, they receive a higher sentence both have payoff of 6. If one prisoner plays D while the other plays C, the player who plays D receives a payoff of 10 while the player who plays C receives the harshest sentence with a payoff of 1. The classical game has the NE of (D,D), even though the Pareto-optimal choice would be (C,C) where both players would do better.

The second two-player game we analyze is the two-player DA's brother game. The payoff matrix for this game is shown below. 
\begin{center}
\begin{tabular}{ l || c | r }
$A| B_1$ & $ \Ket{0}(C)$ &$\Ket{1}(D)$\\
\hline 
\hline
$\Ket{0}(C)$& $ (11,9)$ & $(1,6)$ \\
$\Ket{1}(D)$& $ (10,1)$ & $(6,0)$ 
\end{tabular}
\end{center}

The payoffs for player $A$ are identical to that in the PD game, though player $B$'s are slightly changed. The interpretation of the classical DA brother's game is that player $B_2$'s payoffs change because player $A$ is the DA's brother, and player $B_2$ is afraid that if he remains silent, i.e. plays D, he will get more time in jail (i.e. lower payoff). 

Classically, this game has the NE of (C,C), where players $A$ and $B_2$ receive payoffs 11 and 9 respectively, which are in this case, Pareto-optimal. 

The Bayesian game follows the protocol by Harsanyi \cite{Harsanyi}. The Bayesian game is played between player $A$, and either player $B_1$ or $B_2$, with some probability $p$. This game can be interpreted as player $A$ playing with either player $B_1$ or $B_2$ with the probability $p$, or as a between two players, $A$ and $B$, with incomplete information, parametrized by $p$, where there are two types of player $B$, where type 1 believes $A$ is not the DA's brother, and type 2 believes he is.

The payoff for player $A$ is given by the weighted average of playing with $B_1$ and $B_2$:

\begin{equation}
\langle\$^A(A,B_1) \rangle (p)+ \langle\$^A(A,B_2) \rangle(1-p) 
\label{eq:payoff}
\end{equation}

Whereas the payoff for the $B$ players is given by $\langle \$^{B1}(A,B_1)\rangle$ and $\langle \$^{B2}(A,B_2)\rangle$.

Classically, the Bayesian game has two NE depending on $p$. Player $B_1$ has the dominant strategy D, while player $B_2$'s dominant strategy is C. If player A cooperates, his payoff is $(1 p + 11(1-p))$, but if he plays D, his payoff is $(6 p + 10(1-p))$. If we assume that the $B$ players always play their dominant strategies, then player $A$ will play C if $p<1/6$ and will play D if $p>1/6$. Thus, the NE for $p<1/6$ is (C, D, C) with payoffs $(11-10 p, 10, 9)$ and for $p>1/6$ is (D, D, C) with payoffs $(10-4p, 6,1)$ for players $(A, B_1, B_2)$ respectively.

\section{Quantum Bayesian game theory background}
\label{sec:quantum}

The important aspects of a Bayesian game, namely, facility for cooperation, advice from a referee, the role of nature in introducing partial information, and a constraint to make the game fair, can be made precise in mathematical terms. Quantum probability spaces \cite{KP} are used to describe game situations \cite{Pitowsky} in which players have to choose strategies to maximize their payoffs in a cooperative or non-cooperative manner. 

Let $H$ be a Hilbert space with finite dimension and states are positive operators with unit trace denoted by $ \mathscr{P}$. A quantum probability space is defined by the tuple $(H, P(H), \mathscr{P})$ where $P(H)$ is the set of projections on $H$. In this framework, the expectation of an observable $X$ in the state $\mathscr{P}$ is defined using the trace as $Tr \{\mathscr{P}X\}$ or notationally as $\mathscr{P}(X)$. This notation is justified because a quantum state is a generalization of probabilistic measure. Here, we describe a space for a two-player game with incomplete information to perform Bayesian type of reasoning. In Bayesian games, nature plays a role by assigning a type to each player. 

Let $(\mathscr{C}^2\otimes$$\mathscr{C}^2,\mathbb{A}$$\otimes\mathbb{A}, \mathscr{P}$, $\mathbb{X}$) be a quantum probability space where $\mathscr{P}$ is an entangled state such as $\ket{\Psi_-}=(\ket{0}_A\ket{1}_B - \ket{1}_A\ket{0}_B)/\sqrt{2}$, $\mathscr{A} $ is the *-algebra of Pauli operators, and $\mathbb{X}$ is a set of types each individual player gets from nature. The lack of knowledge on the type received by the other players forces the players to act with incomplete information, thus requiring Bayesian strategies.
It can be shown that when the payoff function is of the following form, closely resembling Bell's inequality, non-classical correlations have an advantage \cite{Noah}:
\begin{equation}
\text{ Cost function} = \mathscr{P}(A_1, B_1) + \mathscr{P}(A_1, B_2) + \mathscr{P}(A_2, B_1) - \mathscr{P}(A_2, B_2), A_i, B_i \in \mathbb{A}. \nonumber
\end{equation}

The non-classical correlation, can be thought of as a piece of advice received by each player from the referee who also makes the final measurements and computes the payoffs, between spatially separated entities. The expectation values of the two players outcomes must respect a no-signaling condition that can be defined as follows:
\begin {equation}
\mathscr{P}(A_1|X_1, X_2) = \sum\limits_{A_2}\mathscr{P}(A_1, A_2|X_1,X_2) = \mathscr{P}(A_1|X_1).
\end{equation}
In other words, in the absence of instantaneous communication, the marginal probability distribution of player A is independent of the
type of player B when we take the types as $X_1 = \text{Type 1}$ and $X_2=\text{Type 2}$. This means the statistics of advice received by player A is independent of type assigned to player B and vice versa, a requirement that would make it a fair game. This condition can be guaranteed by entangling spatially separated states that would prevent instant signaling of the types as part of respecting causality so as not to violate special relativity. In the games considered here the *-algebra is abelian as the observables used in cost functions are compatible whereas the games considered in \cite{Pitowsky} are based on more general non-commutative algebra of operators.

\section{Solution methods}
\label{sec:1}

\begin{figure*}
\includegraphics[width=1\columnwidth]{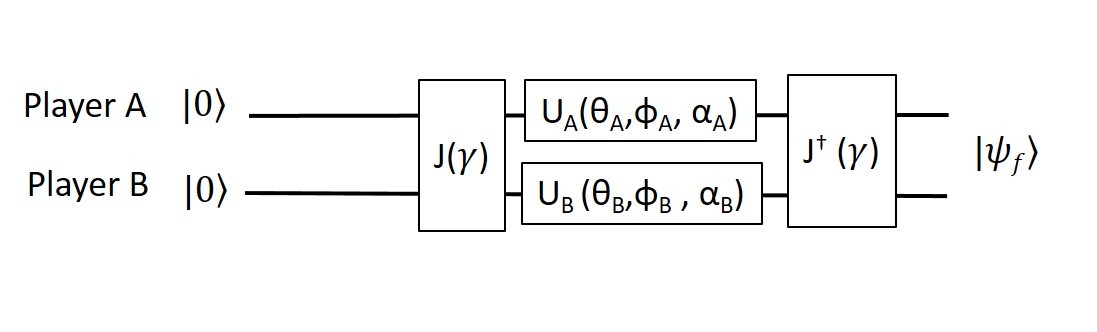}
\caption{\label{fig:QPD}The quantum circuit for the two-player PD game. Both players qubits are initialized to the $\ket{0}$ state, followed by an entangling operation, $J$, that depends on the parameter $\gamma$. Then the players apply their strategy choice, $U_{A,B}(\theta,\phi,\alpha)$, which is followed by an un-entangling operation. The payoffs are determined from the final state $\ket{\psi_f}$.}
\end{figure*}

We analyze a Bayesian game constructed from the two-player quantum game as shown in Fig. \ref{fig:QPD}. We give details of the solution of the two-player games first. We use the EWL quantization scheme\cite{EWL}. In this scheme, qubits that represents the states of individual players are initialized to $\Ket{0}$ followed by an entangling operation $\hat{J}$. Next, each player makes a strategy choice $\hat{U}$. Finally the conjugate transpose of the entangling operation is applied so that if the players do not take any action, the initial state is recovered. At the end of the circuit, the qubits for the two players are measured by the referee and the payoffs are awarded depending on the outcome of the measurement. Because the results of the measurement are probabilistic, all computed payoffs are expectation values. In order to compare the quantum game with the classical version, we can make a correspondence between the outcome $\ket{0}$ and the classical `cooperate' (C) strategy, (i.e. $\hat{U} = \hat{I}_2$, where $\hat{I}_2$ is the identity operator) and $\ket{1}$ with the `defect' (D) strategy (i.e. $\hat{U} = \hat{\sigma}_X$, the Pauli-X operator) and compute the payoff using the tables given in Section \ref{sec:classical}.

In the Bayesian game we analyze, there is one type of player $A$ and two types of player $B$, (i.e. $B_1$ and $B_2$). Player $A$ plays the game shown in Fig. \ref{fig:QPD} with either $ B_1$ or $ B_2$ depending on a probability $p \in [0,1]$. Player $A$'s payoff is the weighted average of playing with $B_1$ and $B_2$, and the $B$ players payoff is normalized by the probability with which they play against player $A$. 

The structure of an un-entangling gate that can be used to define a game has well understood characteristics \cite{Lason}. To compare with other results, we use the commonly found form for our entangling operation:
\begin{equation}
\hat{J}(\gamma)=e^{ i \gamma \hat{\sigma}_x \otimes \hat{\sigma}_x }= 
\begin{pmatrix}
Cos(\gamma/2) & 0 &0 &\imath \: Sin(\gamma/2)\\
0 & Cos(\gamma/2) &-\imath \: Sin(\gamma/2) &0) \\
0 & -\imath \: Sin(\gamma/2) &Cos(\gamma/2) & 0 \\
\imath \:Sin(\gamma/2) & 0 &0 & Cos(\gamma/2)
\end{pmatrix}
\label{eq:J} 
\end{equation}

The parameter $\gamma \in [0,\pi/2]$ defines the amount of entanglement. There is no entanglement when $\gamma = 0$, (i.e. $\hat{J}(0) = \hat{I}_4$), and when $\gamma = \pi/2$, produces a maximally entangled Bell state when operated on the initial state such that $\hat{J}(\pi/2) \cdot \ket{00}= 1/\sqrt{2}(\ket{00} + \imath \ket{11})$.

Though many quantum games are analyzed in a restricted strategy space, it has been pointed out that the solutions to the games are very different if the strategy space is not restricted \cite{Benjamin2001}. The strategies we use are given by a single arbitrary $SU(2)$ rotation of a qubit. This choice allows for any pure strategy. Although the global phase of the two-qubit state is not physical, the relative phase of one qubit with respect to the other is. In order to be fair, we keep the potential strategy choices of the two players symmetric and thus we must specify each player's strategy choices with three parameters $(\theta,\phi,\alpha)$ given by the matrix:

\begin{equation}
\hat{U}(\theta, \phi,\alpha)=
\begin{pmatrix}
e^{-\imath \phi} Cos(\theta/2) & e ^{\imath \alpha} Sin(\theta/2) \\
- e ^{-\imath \alpha} Sin(\theta/2)&e^{\imath \phi} Cos(\theta/2)
\end{pmatrix}
\label{eq:strat}
\end{equation} 

where $\theta \in [0,\pi],\phi \in [0,2\pi],\alpha \in [0,2\pi]$. Because we will employ a numerical solution method, this range of parameters is discretized into a finite number of strategy choices for analysis, as is discussed below.

%%\begin{figure*}
%%\includegraphics[width=1\columnwidth]{bloch.jpg}
%%\caption{\label{fig:bloch}A player's strategy can be visualized as a rotation of vector on Bloch sphere.}
%%\end{figure*}

The outcome of the circuit in Fig. \ref{fig:QPD} is given by: 
\begin{equation}
\ket{\psi_f(A,B)}= \hat{J}^{\dagger}(U_A \otimes U_B)\hat{J}\ket{00}
\label{eq:psif}
\end{equation}

If the payoff for a player $A$, $ \$^A$ is given by a vector in the normal form two-qubit representation (i.e ($\ket{00},\ket{01},\ket{10},\ket{11}$), which is derived from the payoffs in the left side of the bracket in the payoff matrices given in Section \ref{sec:classical}, then the expectation value of the payoff is given by:

\begin{equation}
\langle\$^A(A,B) \rangle = \sum\limits_{j}\Braket{\psi_f(A,B) | \psi_f(A,B)}_j \$^A_j
\label{eq:payoff}
\end{equation}
with analogous expressions for the other players.

To solve for the NE of the game, we use the method of best responses. Analytical solutions have been constructed for the symmetric two-player PD game \cite{Du2003} which allow one to compute the best response to a given strategy choice of an opponent. However, despite our solutions ultimately having a relatively simple representation, an extension of the analytic solution to include asymmetric payoffs and a Bayesian framework with three players remains elusive. In order to have a method of solution that computes all NE of a game, and can easily be used to compare to other payoff matrices (including asymmetric payoffs) in the two-player game and a three-player game, we adopt a numerical approach. Similar to a method that has been used to analyze two-player games and partially analyze a Bayesian game \cite{Avishai2012}. 

We discretize the parameters of the strategy matrix to make list of all possible strategy choices. For example, if step through $\theta, \phi$, and $\alpha$ in steps of $\delta \theta = \delta \phi = \delta \alpha = \pi/8$, the list of strategy choices defines the strategy space, $\mathscr{S}$:
\begin{equation}
\begin{aligned}
\mathscr{S} =&\{\hat{U}(0,0,0),\hat{U}(0,0,\pi/8),\hat{U}(0,0,2\pi/8)... \\
&\hat{U}(\pi,2\pi,6\pi/8), \hat{U}(\pi,2\pi,7\pi/8),\hat{U}(\pi,2\pi,2\pi)\}
\end{aligned} 
\label{eq:strategies}
\end{equation}
Due to the definition of Eq. \ref{eq:strat}, there are several matrices that become redundant because, for example, when $\theta = 0$, $\alpha$ is undefined. 

Next, we construct the best response function for each player, $\mathscr{B}^{A,B} $ with a brute-force method. This is done by computing all of the payoffs for a player against every possible strategy choice of his opponents within $\mathscr{S}$, and selecting the response with the highest payoff. For player $A$, for example, this gives a response function in the form, $\mathscr{B}^A = \{\hat{U}_j^{*A},\hat{U}_j^B\}$ , where $j$ runs over all possible strategy choices in Eq. \ref{eq:strategies} and $\hat{U}_j^{*A}$ is $A$'s best response to $B$'s strategy choice $j$. When this is done for all players, if if the intersection of the best-responses is non-empty, $(\hat{U}_k^{*A},\hat{U}_k^B) =(\hat{U}_k^A,\hat{U}_k^{*B}) $ for some $k$ where $(\hat{U}_k^{*A},\hat{U}_k^B) \in \mathscr{B}^A$ and $(\hat{U}_k^A,\hat{U}_k^{*B}) \in\mathscr{B}^B $, the intersection of the best response curves is a NE.

In analyzing games whose strategy choices are defined by descritazations of Equation \ref{eq:strat}, the numerical analysis becomes impractical. For example, the parameters $(\Delta \theta, \Delta \phi, \Delta \alpha) = (\pi/8, \pi/8, \pi/8)$ yield 1824 unique strategy choices. The computation of all solutions to a two-player game for all values of entanglement for these parameters took nearly an hour, making solutions to the Bayesian game impractical with the current method. We find that the major structures of the games are already captured using the parameters $(\Delta \theta, \Delta \phi, \Delta \alpha) = (\pi, \pi/2, \pi/2)$, which produces a total of eight unique strategy choices. The solutions to these games yielded zero, one, or two unique Nash equilibria, differing by a phase relationship, as will be described below. For games with zero or one Nash equilibrium, the further discretized strategy space yielded no additional solutions. For games with two Nash equilibria, the further discretization of the strategy space resulted in what appeared to be a continuoum of solutions bounded by the two original solutions. The structure and understanding of this continuoum of solutions is the subject of further study that will be reported elsewhere\cite{solmeyerSPIE}. For the remainder of our analysis, the strategy cohices is restricted to the set defined by the stepping parameters $(\Delta \theta, \Delta \phi, \Delta \alpha) = (\pi, \pi/2, \pi/2)$.

\section{Two-player game results}
\label{sec:2}

\begin{figure}
\includegraphics[width=0.75\columnwidth]{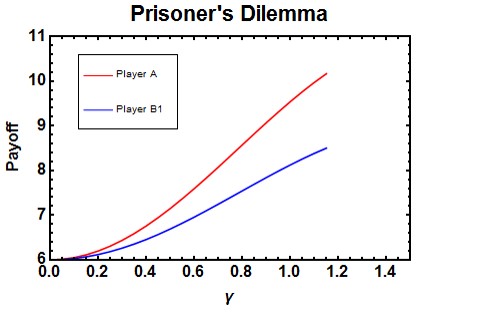}
\caption{\label{fig:PD}Payoff for asymmetric, two-player PD game. The line represents the class of NE with $\theta_A = \theta_{B1} = \pi$. At $\gamma = 0$ the payoff at the classical (D,D) NE is recovered. As the entanglement is increased, the payoff increases until $\gamma = 1.15$, above which there are no Nash equilibrium which is consistent with earlier results that pure strategy NE is absent with maximal entanglement. }
\end{figure}

For the quantum game, we compute the NE as a function of the entanglement parameter $\gamma$ and report the payoff to each player at the NE. As has been seen in previous analyses, there exists a NE for low values of entanglement, and the payoff for both players increases as the entanglement increases, until the entanglement reaches a critical value above which there is no NE. Though these results have been reported before, for completeness, we show the results of our calculation in Fig. \ref{fig:PD}. Our threshold for entanglement is around $\gamma = 1.15$, above which there is no NE. This matches the prediction from the analytic result for the symmetric prisoner's dilemma game \cite{Du2003} if, instead of the asymmetric case we have, we compute the analytic results for payoff for the outcome $\ket{00}$ for both players to be 9.

Each point on the NE plot actually represents a class of NE with an infinite set of strategy choices, all of which have the same payoff. The NE shown in Fig. \ref{fig:PD} has the values of $\theta_A = \theta_{B1} = \pi$. This means that the value of $\phi$ is not defined. The value of $\alpha_A$ can take any value as long as the phase difference between A and B1 takes one of two values, such that $\alpha_{B1} = -\alpha_{A} +\{\pi/2, 3\pi/2\}$. The interpretation of this is the global phase of the two qubit system is not physical, so one of the overall phases is free, while the phase of the other qubit must keep a fixed relationship to the phase of the first. Another way to represent the strategy choice matrices of the NE is as the outer product of Pauli matrices, $\hat{\sigma}_X \otimes \hat{\sigma}_Y$. Though this simple representation does not describe the freedom of the overall phase, it is useful when comparing the quantum strategies to the classical strategies and is added to increase physical intuition. The strategy choices $\hat{\sigma}_X$ and $\hat{\sigma}_Y$ both resemble the classical strategy choice D because they swap the initial state $\ket{0}$ to $\ket{1}$, although their $\alpha$ values differ. The symmetry of the strategy choices in the NE of the quantum game are indicative of the classical NE, and when there is no entanglement, $\gamma = 0$, the game maps onto the classical PD game such that the payoff is identical to the (D,D) NE in the classical PD.

\begin{figure}
\includegraphics[width=0.75\columnwidth]{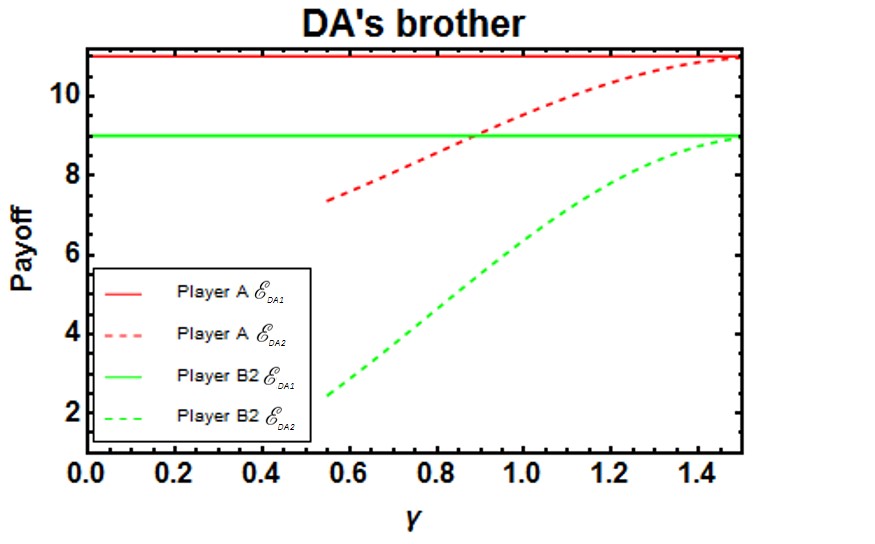}
\caption{\label{fig:DA}Payoff for asymmetric, DA's brother game. The NE labeled $\mathscr{E}_{DA1}$ has the structure $\theta_A = \theta_{B1} =0$, and is independent of the amount of entanglement. The second NE with $\theta_A = \theta_{B1} =\pi$, $\mathscr{E}_{DA2}$, does not exist below $\gamma = 0.55$ and approaches $\mathscr{E}_{DA1}$ as $\gamma$ approaches $\pi/2$. Here pure strategy NE is possible with maximal entanglement due to the fact the NE of the classical game is Pareto efficient.}
\end{figure}
In the quantum version of the DA's brother game, in addition to the multiplicity of solutions in a NE class, the solutions gave two stable NE with different payoffs. In the parametrization of the strategies described in Eqn. \ref{eq:strat}, the first NE ($\mathscr{E}_{DA1}$ of Fig. \ref{fig:DA}) is given by $\theta_A = \theta_{B1} =0$ and phase relationship $\phi_{B2} = -\phi_{A} +\{0, \pi\}$, where $\alpha$ is undefined when $\theta = 0$, and can be represented by the product of operators $\hat{I}_2 \otimes \hat{I}_2$. The NE $\mathscr{E}_{DA1}$ is constant and exists for all values of entanglement, and is equivalent to the classical NE. 

A second NE ($\mathscr{E}_{DA2}$), which has no classical counterpart, appears as the entanglement is increased above $\gamma = 0.55$. This payoff has the structure $\theta_A = \theta_{B1} = \pi$ and phase relationship $\alpha_{B2} = -\alpha_{A} +\{\pi/2, 3\pi/2\}$, where $\phi$ is undefined at $\theta = \pi$. This has the Pauli matrix representation $\hat{\sigma}_X \otimes \hat{\sigma}_Y$. This NE has smaller initial payoffs, but increases towards the (C,C) payoff as the entanglement increases to a maximum of $\gamma = \pi/2$. It is notable that at maximal entanglement, this game has two competing NE with strategy choices differing by more than just a phase relationship, and that have equal payoffs. 

\section{Bayesian game results}
\label{sec:3}

\begin{figure}
\includegraphics[width=1\columnwidth]{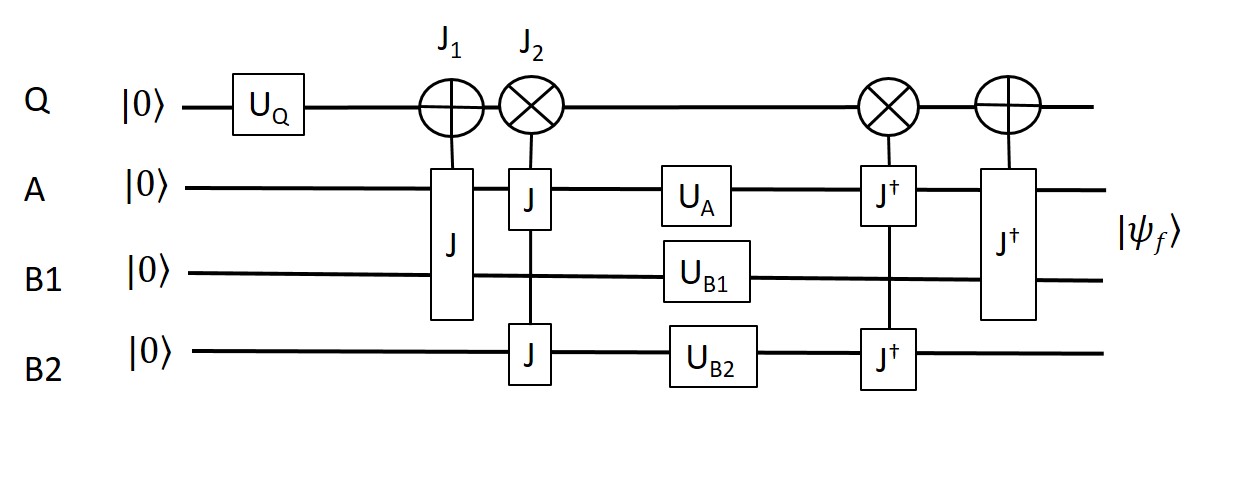}
\caption{\label{fig:quantumcircuit} Quantum circuit for implementing a Bayesian game. $Q$ is the control qubit, $\hat{U}_Q$ defines the probability $p$. The entangling operations are controlled entangling operations such that if the control qubit, $Q$, is in $\ket{0}$, then qubits $A$ and $B_1$ are entangled, where as if $Q$ is $\ket{1}$, then qubits $A$ and $B_2$ are entangled}
\end{figure}

The solutions to the quantum Bayesian game could be computed by probabilistically combining two versions of the two-player circuit shown in Fig \ref{fig:QPD}. However, in order to present a fully quantum formalism, this game can alternatively be encoded in the quantum circuit shown in Fig. \ref{fig:quantumcircuit}. Where the entangling operations are now controlled operations such that A is entangled with $B_1$ or $B_2$ depending on the state of a control qubit $Q$. That is, if $Q$ is $\ket{0}$, then $\hat{J}$ entangles qubits $A$ and $B_1$, where if $Q$ is $\ket{1}$, then $\hat{J}$ entangles $A$ and $B_2$. These are represented in the 3-qubit representation as:

\begin{equation}
\hat{J}_1 = 
\begin{pmatrix}
\hat{I}_2& \hat{0} \\
\hat{0}& \hat{J} 
\end{pmatrix}
\text{ and }
\hat{J}_2 = 
\begin{pmatrix}
\hat{J} & \hat{0} \\
\hat{0}& \hat{I}_2 
\end{pmatrix}
\label{eq:controlledentangle}
\end{equation}

Where $\hat{J}_1$ acts on qubits in the normal representation $(Q, A, B_1)$, $\hat{J}_2$ acts on qubits $(Q, A, B_2)$, $\hat{J}$ is given by Eqn. \ref{eq:J}, and $\hat{0}$ is a 4$\times$4 matrix of zeros.

Allowing $\hat{U}_Q$ to be any arbitrary qubit rotation allows the circuit to realize any value of $p$ such that $p = Sin^2(\theta_Q/2)$ At the end of the game, the control qubit is measured, and the payoff is computed depending on the state of $Q$. A superposition of the control qubit has $A$ play the games with $B_1$ and $B_2$ in parallel.

From the Bayesian game circuit of Fig. \ref{fig:quantumcircuit}, it might appear that using an arbitrary rotation on the control qubit allows us to reach behavior that is not captured by taking a statistical mixture of the games with $B_1$ and $B_2$. However, in practice, we find that the full quantum circuit behaves the same as the statistical mixture, and depends only on the $\theta$, i.e. population, of the control qubit.

\begin{figure}
\includegraphics[width=1\columnwidth]{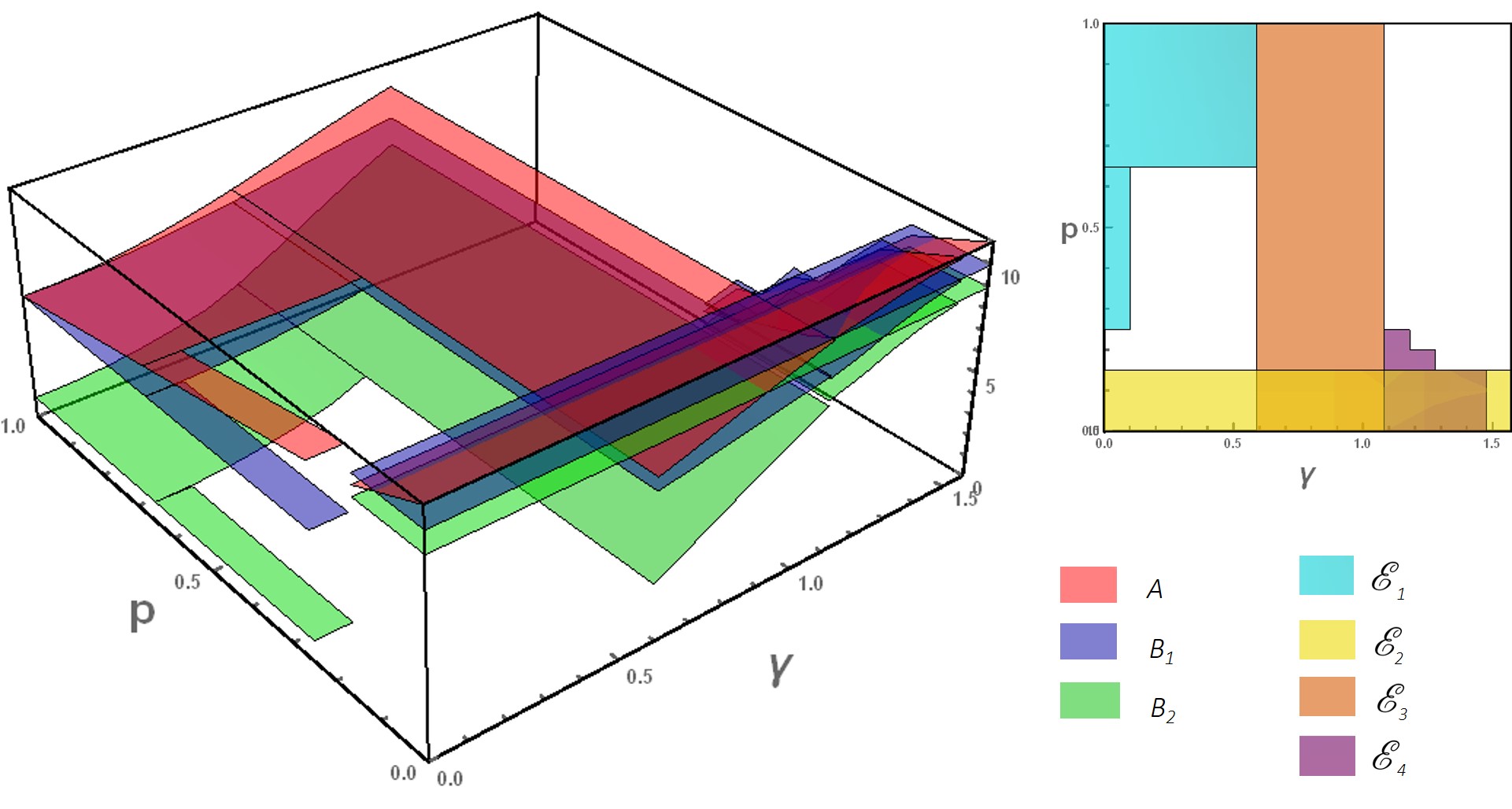}
\caption{\label{fig:Bayesian}Payoff for Bayesian game between $A$, $B_1$ and $B_2$. The red curve represents the payoff at NE for player $A$, the blue curve represents player $B_1$ and the green curve represents $B_2$. Along the axis with $p=1$, the curve is identical to the game in Fig. \ref{fig:PD}, and where $p=0$, it is identical to Fig. \ref{fig:DA}. The above view on the right hand side identifies the regions where $\mathscr{E}_1$-$\mathscr{E}_4$ exist, where they overlap, and where there are no NEs.}
\end{figure}

The results for the quantum Bayesian game are plotted for $\gamma \in [0,\pi/2]$ and $p \in [0,1]$, and are shown in Fig. \ref{fig:Bayesian}. Along the axis with $\gamma = 0$ the results of the quantum game contain the results of the classical game, there are two distinct NE, the payoff of the $B$ players depends only on which NE is being played, and the payoff of player $A$ changes linearly with $p$ for both NE. 

Extending this out into $\gamma \neq 0$ the parameter space is broken up into several regions resembling a phase diagram. Some of the regions have a single NE, some have several, and others have none. The different NE behave differently as functions of $\gamma$ or $p$, as will be described below. As can be seen along the left side of the graph where $P=1$, the result is identical to Fig. \ref{fig:PD}, as $A$ plays solely with $B_1$, and along the right side of the graph where $p=0$, the results are identical to Fig. \ref{fig:DA} as player $A$ plays solely with $B_2$. There are 4 distinct NE in this curve. 

The first NE ($\mathscr{E}_1$) exists in the interval $p \in [1,0.7]$ and $\gamma \in [0,0.55]$ and also exists along the line with $\gamma = 0$ between $p\in[0,0.25] $. This equilibrium has the structure ($\theta_A,\theta_{B1}, \theta_{B2}$) = ($ \pi,\pi,0$) with phase relationships $\alpha_{B1} = \alpha_A - \{\pi/2,3\pi/2\}$ and $\phi_{B2} = \alpha_A - \{\pi/2,3\pi/2\}$, and can be represented by the operators $ \hat{\sigma}_Y \otimes \hat{\sigma}_X \otimes \hat{\sigma}_Z$. 

It should be noted that the operator representations are not unique, because of the possibility of the varying phase relationship between the strategy choices, but they are give more intuition than the parameter representation. The operator representation of NEs for the Bayesian PD game is similar to that of mixed strategies two players quantum PD where the they are shown to have support on Pauli matrices \cite{Landsburg}. The symmetry of the NE strategy choices in the operator representation is suggestive of the (D, D, C) classical NE, as both $ \hat{\sigma}_Y$ and $\hat{\sigma}_X$ have the symmetry of a spin flip which changes the initial cooperate strategy to defect. Similarly, both $\hat{\sigma}_Z$ and the identity operator have the symmetry of the identity operator, which leaves the cooperate strategy unchanged. The relative phase relationships between the player's strategy choices determines which of the operators with the same symmetry are used in the operator representation. 

The payoff at $\mathscr{E}_1$ increases in payoff for the three players as the entanglement is increased. As a function of $p$, player $B_1$ and $B_2$'s payoff are constant while player $A$'s payoff increases linearly as $p$ decreases. It is interesting to note that, in the small values of $\gamma$, the NE disappears below $p=0.7$, long before the second NE takes over as $p<1/6$, leaving a gap where there is no NE. It is also notable that the classical $\gamma=0$ game has a NE for $p \in [.17,0.7]$, whereas for any value of entanglement in the region $\gamma \in (0,.55]$, there is none. 

The constant NE of the two-player DA's brother game forms a second NE (labeled $\mathscr{E}_2$) and exists for small $p$ values, but ceases to exist for $\sim p>0.15$. This is intuitive since player $A$ is mostly playing with player $B_2$, so their behavior dominates the structure of the equilibria, with player $B_1$ playing his best response to their strategies at equilibria, which notably, is the highest payoff of all players in the region of $p \in [0.07,0.15]$. The payoff of this NE, and its extent along the $p$ axis are both independent of $\gamma$. Player $B_1$ and $B_2$'s payoffs are constant at 10 and 9 respectively. $A$'s payoff decreases linearly as $p$ is increased from 11 at $p=0$ to 9 at $p \sim 0.15$. $\mathscr{E}_1$ has the structure ($\theta_A,\theta_{B1}, \theta_{B2}$) = ($ 0,\pi,0$). It has the phase relationships given by $\alpha_{B1} = \phi_A - \{0,\pi\}$ and $\phi_{B2} = \phi_A - \{0,\pi\}$. $\mathscr{E}_2$ can be represented by the operators $\hat{I}_2 \otimes \hat{\sigma}_Y \otimes \hat{I}_2$, which is suggestive of the classical NE (C, D, C). 

When $\gamma > 0.5$, a third NE ($\mathscr{E}_3$) occurs with the structure ($\theta_A,\theta_{B1}, \theta_{B2}$) = ($ \pi,\pi,\pi$) with phase relationships $\alpha_{B1} = \alpha_A - \{\pi/2,3\pi/2\}$ and $\alpha_{B2} = \alpha_A - \{\pi/2,3\pi/2\}$, and can be represented by the operators $ \hat{\sigma}_Y \otimes \hat{\sigma}_X \otimes \hat{\sigma}_X$. $\mathscr{E}_3$ looks in payoff like it is the continuation of $\mathscr{E}_2$ in the region $p \in [1,0.7]$, only with player $B_2$ changing his strategy. Though the intuitive interpretation of player $A$ and player $B_1$'s strategy choices map onto their classical two-player NE (D, D), since player $B_2$ changes his strategy from $\mathscr{E}_1$, this represents a NE that doesn't occur in the classical version of the game. There are more NEs of type $\mathscr{E}_3$ than others, as the phase relationship between the strategies is the most flexible indicating that the more opportunities to cooperate, the larger then number of equilibria.

A fourth NE ($\mathscr{E}_4$) exists with the structure $(\theta_A,\theta_{B1}, \theta_{B2}) = (\pi,0,\pi)$ with phase relationships $\phi_{B1} = \alpha_A - \{\pi/2,3\pi/2\}$ and $\alpha_{B2} = \alpha_A - \{\pi/2,3\pi/2\}$, and can be represented by the operators $ \hat{\sigma}_Y \otimes \hat{\sigma}_Z \otimes \hat{\sigma}_X$. $\mathscr{E}_4$ has a small range of $p\sim \in [0,0.2]$ and $\gamma\sim \in [1.2,1.45]$. There is a small region of parameter space near $p \sim .2$ and $\gamma \sim 1.2$, where this is the only NE. In this small region, player $B_1$ has the highest payoff, followed by $A$ and then by $B_2$. 

The results describing the parameters of the strategies of the various NE are summarized in Tables \ref{tab:table1} and the range as well as operator representations are summarized in Table \ref{tab:table2}.
\begin{table}

\begin{tabular}{l|l|l|l|l|l}

& $\{\theta_A,\theta_{B1},\theta_{B2}\}$ & $\{\phi_A,\phi_{B1},\phi_{B2}\}$ & $\{\alpha_A,\alpha_{B1},\alpha_{B2}\}$ \\
\hline

$\mathscr{E}_1$ &$\{\pi,\pi,0\}$& $\{\emptyset,\emptyset,X-\{\pi/2,3\pi/2\}\}$ & $\{X,X-\{\pi/2,3\pi/2\},\emptyset\}$ \\ 
$\mathscr{E}_2$ &$\{0,\pi,0\}$ & $\{X,\emptyset,X-\{0,\pi\}\}$ & $\{\emptyset,X-\{0,\pi\},\emptyset\}$ \\ 
$\mathscr{E}_3$ &$\{\pi,\pi,\pi\}$ & $\{\emptyset,\emptyset,\emptyset\}$ &$\{X,X-\{\pi/2,3\pi/2\},X-\{\pi/2,3\pi/2\}\}$ \\ 
$\mathscr{E}_4$ &$\{\pi,0,\pi\} $& $\{\emptyset,X-\{\pi/2,3\pi/2\},\emptyset\}$ & $\{X,\emptyset,X-\{\pi/2,3\pi/2\}\}$ 

\end{tabular}

\caption{\label{tab:table1}Summary of the strategy parameters of the NE of the three-person Bayesian game. $\emptyset$ means that the parameter is undefined. $X$ means that the parameter can take any value, so long as the parameters of the other players obey a given phase relationship. }
\end{table}

\begin{table}
\begin{tabular}{l|l|l|l}
& Range $(p)$ & Range $(\gamma)$ &Operator representation \\
\hline
$\mathscr{E}_1$ &$[0.7, 1]$ & $[0, 0.55]$ & $ \hat{\sigma}_Y \otimes \hat{\sigma}_X \otimes \hat{\sigma}_Z$ \\ 
$\mathscr{E}_1^*$ &$[0, 0.25]$ & $0$ & $ \hat{\sigma}_Y \otimes \hat{\sigma}_X \otimes \hat{\sigma}_Z$ \\
$\mathscr{E}_2$ & $[0, 0.15]$ &$[0, \pi/2]$ & $ \hat{I}_2 \otimes \hat{\sigma}_Y \otimes \hat{I}_2$ \\ 
$\mathscr{E}_3$ &$[0, 1]$ &$[0.55, 1.1]$ & $ \hat{\sigma}_Y \otimes \hat{\sigma}_X \otimes \hat{\sigma}_X$ \\ 
$\mathscr{E}_4$ &$[0, 0.2]$ &$[1.2, 1.45]$ & $ \hat{\sigma}_Y \otimes \hat{\sigma}_Z \otimes \hat{\sigma}_X$

\end{tabular}
\caption{\label{tab:table2}Summary of the ranges of $\mathscr{E}_1$-$\mathscr{E}_4$ and their operator representation. The operator representation is not unique, and is only one of the possible strategy choices of player A and one of the possible phase relationships to player B}
\end{table}

In addition, there are two blocks of parameter space where there are no NE. They are given by $p \in [0.2,1.0] $ and $\gamma \in [\pi/2,1.15] $ and in the region $p \in [0.2,0.7] $ and $\gamma \in (0,0.55] $.

\section{Discussion}
\label{sec:4}

The interpretation of a NE is that by playing rationally, the players in a game will tend towards playing the NE strategy choices. The NE is stable in that players do not have any incentive to deviate, and is thus self-enforcing. The character of the equilibria that arise when entanglement is present seem to bear resemblance to the concept in classical games of a correlated equilibrium\cite{Auman1974}. A correlated equilibrium in classical games is achieved when mixed strategies are employed and there is communication between the players in the form of advice or a contract. If players receive some piece of advice, or react in a predetermined way to a random event, they can employ strategies that are correlated with one another and realize self-enforcing equilibria that are different from those in the mixed game without communication. 

In contrast, our analysis includes only pure strategies and the role of the advice is played by the initial entanglement. When the player's qubits are entangled, the outcomes of the measurement following their strategy choices will be correlated in a specific way, determined by the type of entanglement. The entanglement is imposed on the players by a referee, and once it is initially established, no communication between the players is necessary, and in fact the players are physically prohibited from breaking the contract. The correlation will also persist even if the players make their strategy choices simultaneously and non-locally. In addition, the specific type of non-classical correlation, enabled by entanglement, can be such that the players can have correlated outputs that are not possible with classical probability distributions in the absence of communication.

However, as the amount of entanglement is changed, the effect that the imposed correlation has on the structure of the NE can change dramatically. At zero entanglement, the quantum formulation of the game strongly resembles the classical game, with the players playing quantum strategies that closely resemble the classical strategy choices (D,D). As the entanglement increases, the strategy choices of the NE do not change, rather, they continue to play the same strategies as in the case with no entanglement. Thus, the contract enforced by the initial entanglement does not induce the players to play a strategy that is different than the one they would play if there were no entanglement, rather it ensures that the outcome of their strategy choices is correlated in a certain way. The conditions of the NE guarantee that the resulting equilibrium is self-enforcing. 

As the degree of entanglement, or amount of correlation, is increased, one might expect that the effective contract is more strictly enforced or that the advice is more closely followed, leading to a larger benefit for the players at equilibrium. This is true for a while, but as in the two-player quantum PD game with symmetric payoffs, the NEs are absent above a critical value of entanglement. This is similar to the phase-transition like behavior that has been seen in some quantum games \cite{Du2003}, and should be investigated further. This also contrasts with an intuition of a classical game with correlated equilibrium, where one might expect that the more strictly a contract is enforced, the greater the benefit from that contract. 

In the Bayesian game, we see evidence of a structure with a much richer and sometimes surprising phase-transition like behavior that can occur both in the amount of entanglement, as has been seen in the two-player games, but also in the amount of incomplete information, i.e. $p$. If only looking at the classical Bayesian game, and the two versions of the two-player game, one would not necessarily predict that there is a region in the center with no NE, that new equilibria may appear (i.e. $\mathscr{E}_4$), or how each of the NE will depend on the parameters $p$ and $\gamma$ without solving for the full possibilities of the game. It is perhaps indicative of the structure of classical probability theory and quantum mechanical probability theory that the payoffs at the NE vary linearly along $p$ the classical probability and non-linearly (i.e. as trigonometric functions) along the quantum parameter $\gamma$, in which probabilities depend on the square of the components of the player's state (i.e. wave function).

The fact that the amount of entanglement can produce abrupt changes in the behavior of a quantum game underscores the importance of decoherence in a quantum game application, as the purity of the initially entangled state could dramatically influence the outcomes and the stability of the game. The behavior of the game can also abruptly changes as a function of the agents prior knowledge in a game, i.e. $p$. This could certainly impact any algorithm taking place on a network, where knowledge of the motivations and abilities of the other agents on the network is incomplete. The structure as a result of the amount of entanglement is due to the constraints imposed by the referee, while the structure in the priors is dependent on the beliefs of the players. Both are critical to the structure of the game, however, the referee can constrain the possible equilibria that may be achieved by adjusting the amount of entanglement, even if the player's preferences, i.e. payoffs, and prior knowledge remain unchanged.

As stated earlier, each NE solution in the quantum game is an infinite class of equilibria with a fixed phase relationship and equal payoffs. There remains an uncertainty in exactly which equal payoff NE, i.e. which phase, the players will end up playing. This uncertainty can also arise in classical games. In several cases, there are multiple NE that are different payoffs, and that differ by more than just a phase relationship, such as in the DA's brother game. The $\mathscr{E}_4$ equilibrium in the Bayesian game is an example of a NE that does not exist in the two-player games, and can occur simultaneously with other NE. When there are multiple NE, while playing a game, it would be possible for the players to be stuck on a lower payoff NE, as a local maximum in their payoff landscape, neither being willing to deviate.

It is interesting to note that in the two-player DA's brother game, or with the corresponding NE in the Bayesian games, the multiple NE that exist have strategy choices which differ by more than a phase relationship and even correspond to different classical strategy choices. At lower entanglement they have distinct payoffs, but as the entanglement approaches maximal, the payoff converges to the same value. This could complicate the NE when the two players cannot agree on which of the equivalent payoff NE to play.

\section {Conclusions and Future Work}
We have classified the solutions to a quantum Bayesian game based on the prisoner's dilemma where there are multiple Nash equilibria. The phase structure of the game in entanglement and probability space is non-trivial. The payoffs at these Nash equilibria are dependent on the entanglement parameter and the probability to play with either player, and we have also identified some regions with the absence of Nash equilibria. We solved for the phase relationships between the sets of strategy choices within each class of Nash equilibria. We have seen evidence of a phase-transition like behavior of the quantum Bayesian game varying both with the amount of entanglement and the degree of incomplete information.

The relationship of the equilibrium solutions produced in the quantum game with entanglement to the correlated equilibrium in classical games should be explored further. The role of entanglement is often interpreted as a type of communication or contract, yet the correlations induced by entanglement persist even when communication is not allowed. Entanglement is one of the more powerful and interesting properties of quantum mechanics and a referee may be able to expoit the effective non-local contract it forms in applications on a quantum network. 

The phase-like behavior of the quantum games should be investigated further to determine the nature of the phase transitions that occur. The player's beliefs, or their priors, can result in entirely different equilibria forming and the role of entanglement as a contract could potentially be elucidated by better understanding how the amount and type of entanglement, and the player's prior beliefs can lead to phase transition-like structures in the Nash equilibria of a game. 

\bibliographystyle{abbrvnat}

\end{document}